\documentstyle[aps,epsfig,twocolumn]{revtex}
\begin{document}
\draft
\flushbottom
\twocolumn[
\hsize\textwidth\columnwidth\hsize\csname @twocolumnfalse\endcsname

\title{
Discontinuous transitions in double exchange materials}
\author{J. L. Alonso$^1$, L. A. Fern\'andez$^2$, F. Guinea$^3$,
V. Laliena$^1$ and V. Mart{\'\i}n-Mayor$^4$}
\address{
$^1$ Dep. de F{\'\i}sica Te\'orica, Facultad de Ciencias,
Universidad de Zaragoza, 50009 Zaragoza, Spain.\\
$^2$ Dep. de F{\'\i}sica Te\'orica, Facultad de CC. F{\'\i}sicas,
Universidad Complutense de Madrid, 28040 Madrid, Spain.\\
$^3$ 
Instituto de Ciencia de Materiales (CSIC). Cantoblanco,
28049 Madrid. Spain. \\
$^4$ Dip. di Fisica,
Universit\`a di Roma ``La Sapienza'', 
Ple. Aldo Moro 2, 00185 Roma and
INFN sezione di Roma, Italy.}
\date{\today}
\maketitle 
\tightenlines
\widetext
\advance\leftskip by 57pt
\advance\rightskip by 57pt

\begin{abstract}
It is shown that the double exchange Hamiltonian, with weak antiferromagnetic
interactions, has a rich variety of first order transitions between
phases with different electronic densities and/or magnetizations.
For band fillings in the range $0.3 \le x \le 0.5$, and at finite
temperatures, a discontinuous PM-FM
transition between phases with similar electronic densities but 
different magnetizations takes place.
This sharp transition, which is
not suppressed by electrostatic effects, and survives in the presence
of an applied field, is consistent with the phenomenology
of the doped manganites near the transition temperature.
\end{abstract}

]
\narrowtext
\tightenlines
\section{Introduction.}
Doped manganites show many unusual features, the most striking being
the colossal magnetoresistance (CMR) 
in the ferromagnetic (FM) phase~\cite{WK55,KS97,CVM99}.
In addition, the manganites have a rich phase diagram as function
of band filling, temperature and chemical composition. The broad
features of these phase diagrams can be understood in terms of
the double exchange model (DEM)\cite{Z51,AH55}, 
although Jahn-Teller deformations\cite{MLS95}
and orbital degeneracy may also play a role\cite{BKK99}.
A remarkable property of these compounds is the existence of
inhomogeneities in the spin and charge distributions
in a large range of dopings, compositions
 and temperatures\cite{Tetal97,Uetal99,Fetal99}.
At band fillings where CMR effects are present, $x \sim 0.2 - 0.5$,
these compounds can be broadly classified into those with a high
Curie temperature and a metallic paramagnetic phase, and those
with lower Curie temperatures 
and an insulating magnetic phase\cite{Fetal96,Fetal98,Metal99}.

The DEM is a simplification of the FM
Kondo lattice, where the FM coupling between core spins
and conduction electrons is due to Hund's rule. When this coupling is
larger than the width of the conduction band, the model
can be reduced to the double
exchange model with weak inter-atomic antiferromagnetic (AFM) interactions.
Early investigations\cite{G60} showed a rich phase diagram, with
AFM, canted and FM phases, depending on
doping and the strength of the AFM couplings.
More recent studies have shown that the competition between 
the double exchange and the AFM couplings leads
to phase separation into AFM and FM regions,
suppressing the existence of canted phases\cite{N96,RHD97,AG98,Yetal98}.
In addition, the double exchange mechanism alone induces a change in the
order of the FM transition, which becomes of first order,
and leads to phase separation, at low dopings\cite{AGG99}.
Note, however, that a detailed study of the nature of
the transition at finite temperatures is still lacking,
despite its obvious relevance to the experiments.

The purpose of this work is to investigate systematically the phase
diagram of the DEM with weak AFM 
interactions.
 We find, in addition to the previously discussed transitions,
a PM-FM first order transition near half filling, if the double exchange
mechanism is sufficiently reduced by the AFM 
interactions.  This transition does not involve a significant change
in electronic density, 
so that domain formation is not suppressed by electrostatic effects.

The model is described in the next section, and the method of
calculation is introduced in the following section. The main results
are presented in section IV, and the main conclusions are discussed
in section V.
\section{The model.}
We study a cubic lattice with one orbital per site. At each site there is
also a classical spin. The coupling between the conduction electron and this 
spin is assumed to be infinite, so that the electronic state with spin
antiparallel to the core spin can be neglected. Finally,
we include an AFM coupling between nearest neighbor core
spins~\cite{FOOTNOTE1}. The Hamiltonian is:
\begin{equation}
{\cal H} =\sum_{ij} 
 {\cal T}( \mbox{\boldmath$S$}_i,\mbox{\boldmath$S$}_j )
c_i^\dag c_j +
\sum_{\langle i,j\rangle } \tilde{J}_{\mathrm AF} 
S^2\mbox{\boldmath$S$}_i\cdot\mbox{\boldmath$S$}_j
\label{hamil}
\end{equation}
where $S = 3/2$ is the value of the spin of the core,
Mn$^{3+}$, and $\mbox{\boldmath$S$}$ stands for a unit vector oriented parallel
to the core spin,
which we assume to be classical. In the following,
we will use $J_{\mathrm AF} = \tilde{J}_{\mathrm AF} S^2$. 
Calculations show that
the quantum nature of the core spins 
does not induce significant effects\cite{AG98}. 
The function
${\cal T}( \mbox{\boldmath$S$}_i,\mbox{\boldmath$S$}_j )=
t\,[\cos\frac{\theta_i}{2}\cos\frac{\theta_j}{2}+
\sin\frac{\theta_i}{2}\sin\frac{\theta_j}{2}
{\mathrm e}^{{\mathrm i}(\varphi_i-\varphi_j)}]$
stands for the overlap of
two spin 1/2 spinors oriented along the directions defined by
$\mbox{\boldmath$S$}_i$ and $\mbox{\boldmath$S$}_j$, whose polar and azimuthal
angles are denoted by $\theta$ and $\varphi$, respectively. 
We study materials of composition
La$_{1-x}M_x$MnO$_3$, where $M$ is a divalent ion, and
$x \le 0.5$. In this composition range, the probability
of finding two carriers in neighboring sites (two 
contiguous Mn$^{4+}$ ions) is small, so that a carrier in a given
ion has all the $e_g$ orbitals in the next ions available. Then,
the anisotropies associated to the differences between the
two inequivalent $e_g$ orbitals should not play a major role.
On the other hand, if $x \ge 0.5$, we expect a significant dependence
of the hopping elements on the occupancy of orbitals
in the nearest ions. In this regime, the equivalence of the two $e_g$
orbitals in a cubic lattice can be broken, leading to
orbital ordering\cite{BKK99,Hetal00} (see, however\cite{K00}).
We will show that the main features of the PM-FM phase
transition, for $x \le 0.5$, can be understood without
including orbital ordering effects. Moreover, 
in this doping range, anisotropic manganites show similar
features\cite{Hetal98,Metal99b,VDetal99,GLetal00}, which suggest
the existence of a common description for the transition.
We will also neglect the
coupling to the lattice. 
As mentioned below, magnetic couplings
suffice to describe a number of discontinuous transitions
in the regime where CMR effects are observed.
These transitions modify substantially the coupling between
the conduction electrons and the magnetic excitations.
Thus, they offer a simple explanation for the anomalous
transport properties of these compounds. 
Couplings to additional
modes, like optical or acoustical phonons \cite{Ietal95},
and dynamical Jahn-Teller distortions \cite{Betal98}
will enhance further the
tendency towards first order phase transitions discussed here.
 
We consider that
a detailed understanding of the role of the magnetic interactions
is required before adding more complexity to the model.

\section{Method.}

At finite temperatures, the thermal disorder in the orientation of
the core spins induces off-diagonal disorder in the dynamics of the
conduction electrons. The calculation of the 
partition function requires an average over core spin textures,
weighted by a Boltzmann factor which depends on the
energy of the conduction electrons propagating within each texture.
We have simplified this calculation by replacing 
the distribution of spin textures
by the one induced by an effective field acting on the
core spins, which is optimized
so as to minimize the free energy. The electronic energy
includes accurately the effects of the core spin disorder on the 
electrons. Our calculation is a mean field approximation 
to the thermal fluctuations of the core spins, retaining, however, 
the complexity of a system of electrons with off-diagonal disorder.
This approximation can be justified by noting that the conduction electrons
induce long range interactions between the core spins, that always
favor a FM ground state. In general, our method is well
suited for problems of electrons interacting with classical fields.

In more mathematical terms, we have used the variational formulation
of the Weiss Mean-Field method~\cite{PARISI} to compute the free energy of
the system. We first trace-out
the fermion operators in (\ref{hamil}), thus obtaining the effective
Hamiltonian for the spins, 
\begin{eqnarray}
{\cal H}^{\mathrm{eff}}(\{\mbox{\boldmath$S$}\})&=& 
J_{\mathrm AF} 
\sum_{\langle i,j\rangle }\mbox{\boldmath$S$}_i\cdot\mbox{\boldmath$S$}_j 
-\\\nonumber
&-&k_{\mathrm B}TV\int d E\,
g(E;\{\mbox{\boldmath$S$}\})\,
\log\left[1+{\mathrm e}^{-\frac{E-\mu}{k_{\mathrm B}T}}\right]\,,
\label{ELIBRE_F}
\end{eqnarray}
where $g(E;\{\mbox{\boldmath$S$}\})$, is the fermionic density 
of states and $V$ the volume of the system.
The Mean-Field procedure consists
on comparing the system under study with a set of simpler reference models,
whose Hamiltonian ${\cal H}_0$ depends on external parameters.
We choose
\begin{equation}
{\cal H}_0= -\sum_i \mbox{\boldmath$h$}_i\cdot\mbox{\boldmath$S$}_i\,.
\label{H0}
\end{equation}
The variational method follows from the inequality
\begin{equation}
{\cal F}\leq {\cal F}_0 + \langle {\cal H}_{\mathrm eff} - {\cal H}_0
\rangle_0\,,
\label{VAREQ}
\end{equation}
where ${\cal F}_0$ is the free energy of the system with
Hamiltonian (\ref{H0}), and 
the expectation values $\langle\cdot\rangle_0$
are calculated with the Hamiltonian ${\cal H}_0$. 
The {\em mean-fields} $\{\mbox{\boldmath$h$}\}$ are chosen to minimize 
the right-hand side of (\ref{VAREQ}). The calculation of the right-hand
side of (\ref{VAREQ}), requires the average of the density
of states (see Eq.(\ref{ELIBRE_F})) on spin configurations
straightforwardly generated according to the Boltzmann weight 
associated to the Hamiltonian ${\cal H}_0$ and temperature $T$.
The key point is that 
$g(E;\{\mbox{\boldmath$S$}\})$  can be numerically calculated
on very large lattices without further
approximations using the method of moments~\cite{MOMENTI} 
(complemented with an standard truncation procedure~\cite{TRUNCACION}).
We have extracted the spin-averaged density of states on a $64 \times 64
\times 64$ lattice (for these sizes, we estimate that finite size
effects are negligible). For simplicity on the
analysis, we have restricted ourselves to four families of fields:
uniform, $\mbox{\boldmath$h$}_i=\mbox{\boldmath$h$}$, giving rise to
FM ordered textures; $\mbox{\boldmath$h$}_i=(-1)^{z_i}\mbox{\boldmath$h$}$,
originating A-AFM order, i.e., textures that are FM within planes
and AFM between planes; $\mbox{\boldmath$h$}_i=(-1)^{x_i+y_i}\mbox{\boldmath$h$}$,
producing C-AFM order, that is, textures that are FM within lines and
AFM between lines; and staggered, 
$\mbox{\boldmath$h$}_i=(-1)^{x_i+y_i+z_i}\mbox{\boldmath$h$}$, which 
originate G-AFM order, i.e., completely AFM textures.
We have chosen fields of these kind since they produce the expected kinds 
of order, although this is not a limitation of the method. 
Once the spin-averaged density of states is
obtained, it is straightforward to obtain the values of the mean-field
that minimize the right-hand size of Eq.(\ref{VAREQ}), and the
corresponding value of the density of fermions.
Expressing the right-hand side of Eq.(\ref{VAREQ}) as a function of the
magnetization (or staggered magnetizations), we obtain the
Landau's expansion of the free energy on the order parameter.

It is finally worth mentioning when our calculation and the Dynamical
Mean Field Approximation~\cite{Yetal98,FURUKAWA} are expected to yield
the same results. It is clear that the key point is the calculation of
the density of states in Eq.(\ref{ELIBRE_F}).  For this problem of
classical variables, the dynamical Mean-Field is known to yield the
same density of states than the CPA
approximation~\cite{GEORGES}. Under the hypothesis of spatially
uncorrelated fluctuations of the spins, which holds in any Mean-Field
approximation, the CPA becomes exact on the Bethe lattice with large
coordination number. However, one cannot conclude that with our 
calculation we would get the same results on the Bethe lattice, since
one has still to specify the probability distribution for the
spins to be used in the CPA calculation of the average density of
states. In Refs.~\cite{FURUKAWA,Yetal98} the calculation is done
by identifying an effective Heisenberg-like mean field, which
becomes exact when the magnetization is very small. Then,
the distribution of spin orientations is equivalent to
the one generated by
an effective magnetic field. In this limit, our ansatz
should reproduce the calculations reported in\cite{FURUKAWA,Yetal98},
when implemented in a Bethe lattice. 

In order to study first order transitions, one must consider solutions
at finite magnetizations. Then, the optimal Boltzmann weights need not
coincide with the effective field ansatz made here.  Detailed DMFA
calculations for the double exchange model\cite{GGA00,G00}, however,
show that the differences between the optimal DMFA distribution and
that obtained with an effective field are small throughout the entire
range of magnetizations.  Thus, the scheme used in this work includes
the same physical processes as the DMFA, but it is also able to
describe effects related to the topology of the three-dimensional
lattice, like those associated to the Berry phase, which arises from
the existence of closed loops. Furthermore, the present scheme allows
us to study the relative stability of phases, like the A and C
antiferromagnetic phases described below, which can only be defined in
a cubic lattice.

\section{Results.}

\begin{figure}
\centerline{\epsfig{file=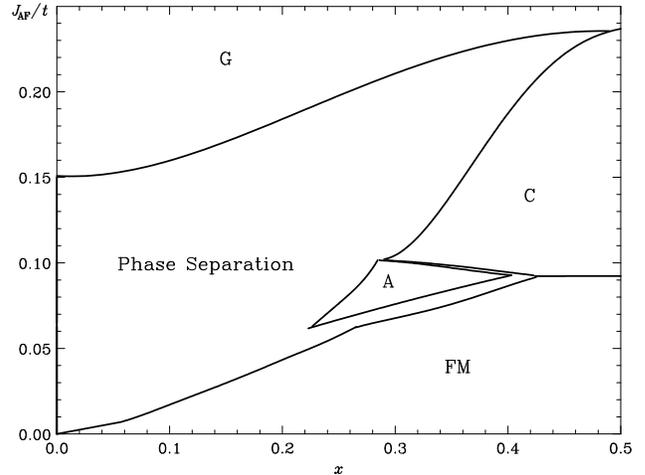,width=2.5in,angle=90}}
\caption{Calculated phase diagram at $T=0$.
The A-AFM phase has ferromagnetic alignment within planes, and
antiferromagnetic alignment between parallel planes.
The C-AFM phase has  ferromagnetic alignment along chains,
and antiferromagnetic alignment between neighboring chains.}
\label{fig:T0}
\end{figure}

The model, Eq.(\ref{hamil}), contains two dimensionless parameters,
the doping $x$, and the ratio $J_{\mathrm AF} / t$. The range of
values of $x$ is $0 \le x \le 1$, and the Hamiltonian has
electron-hole symmetry around $x = 0.5$. The zero temperature phase
diagram, shown in Fig.~\ref{fig:T0}, is calculated minimizing the
effective Hamiltonian at fixed chemical potential and zero temperature
(we take the limit of zero temperature in Eq.(\ref{ELIBRE_F}) obtaining the
grand-canonical Hamiltonian), within the four Mean-Field ansatzs
previously defined. At zero $J_{\mathrm AF} / t$, only the
ferromagnetic phase is found, and the system is stable at all
compositions. When $J_{\mathrm AF} / t$ is finite, there is a value of
the chemical potential for which the empty system with a perfect
G-type AFM spin ordering has the same {\em grand-canonical} energy
that a system with a perfect FM spin ordering and a finite value of
$x$. At this value of the chemical potential the system is unstable
against phase separation~\cite{AGG99}, as shown in
Fig.~\ref{fig:T0}. Notice that the phase-separation region can never
reach $x=0.5$, due to the hole-particle symmetry. For larger values of
$J_{\mathrm AF}$ a small region of A-type AFM is found for
$x\sim0.25$, and a much larger region of C-type AFM for $x$ close to
half-filling. Finally, a G-type AFM-region is eventually reached by
further increasing $J_{\mathrm AF}/t$. However, this is not a
saturated antiferromagnetic phase since the mean-field that minimizes
the grand-canonical energy has a finite $\mbox{\boldmath$h$}/T$ when
$T$ tends to zero~\cite{FOOTNOTE2} (notice that one cannot have a
continuously varying value of $x$ in a perfect AFM configuration).

\begin{figure}
\centerline{\epsfig{file=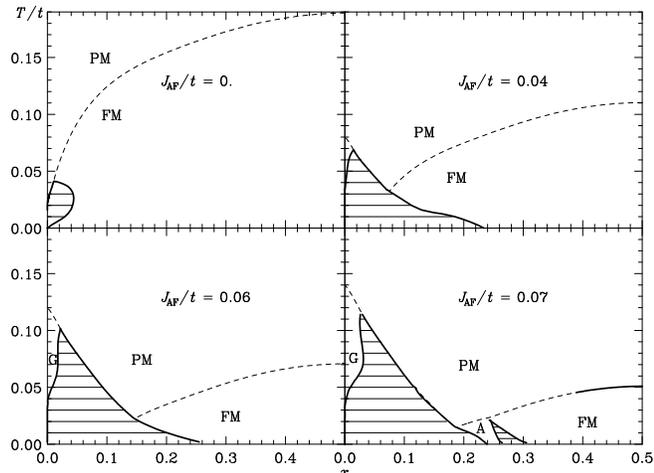,width=2.5in,angle=90}}
\caption{ Transition temperatures as function of
electronic density and strength of the AFM couplings.
The dashed lines correspond to continuous transitions. Solid thick
lines are drawn for first order transitions, and the stripes
correspond to phase coexistence regions. 
The onset of first order transitions
at $x\sim 0.5$ is $J_{\mathrm AF}/t\approx 0.06$.}
\label{fig:phased}
\end{figure}

Let us now discuss the phase diagrams at non-zero temperatures for the
different values of $J_{\mathrm AF} / t$  shown in
Fig.~\ref{fig:phased}.  For $J_{\mathrm AF} = 0$, we obtain a maximum
transition temperature of $T = 400$K for a width of the conduction
band $W = 12 t \approx 2$eV, which is consistent with a density of
states of $\rho ( E_{\mathrm F} ) = 0.85$ eV$^{-1}$ calculated
in\cite{PS96} for La$_{1/3}$Ca$_{2/3}$MnO$_3$ (see
also\cite{Getal96}).  Note that the bandwidth calculated in this way
is probably an overestimate, as it does not include renormalization
effects due to lattice vibrations\cite{Letal97}.  There is some
controversy regarding the value of $J_{\mathrm AF}$.  The reported
value of $J_{\mathrm AF}$ for the undoped compound, LaMnO$_3$, is
$\tilde{J}_{\mathrm AF} \approx 0.58$meV, so that $J_{\mathrm AF}
\approx 0.005 t$\cite{Metal96}, although calculations give higher
values\cite{PGJ98}.  In the doped compounds, there is an additional
contribution of order $J_{\mathrm AF} \sim t^2 / U_{\mathrm ex}$,
where $U_{\mathrm ex} \approx 1 - 2$eV is the level splitting induced
by the intra-atomic Hund's coupling\cite{CVM99}.  Thus, $J_{\mathrm
AF} \sim 0.01 t - 0.08 t$, although higher values have been
suggested\cite{BKK99}.

Our results show four types of first order transitions:

i) In pure DEM ($J_{\mathrm AF} = 0$) the magnetic 
transition becomes discontinuous at sufficiently low densities, in agreement 
with the analysis presented in\cite{AGG99}. The phase coexistence region
shrinks to zero and the critical temperatures vanish as $x$ goes to zero,
as expected.

ii) The competition between antiferromagnetism and ferromagnetism
when $J_{\mathrm AF} \ne 0$ leads to a discontinuous transition which 
prevents the formation of canted phases, as reported in\cite{N96,RHD97,AG98}.
This transition also takes place at low dopings.

\begin{figure}
\begin{raggedleft}
\epsfig{file=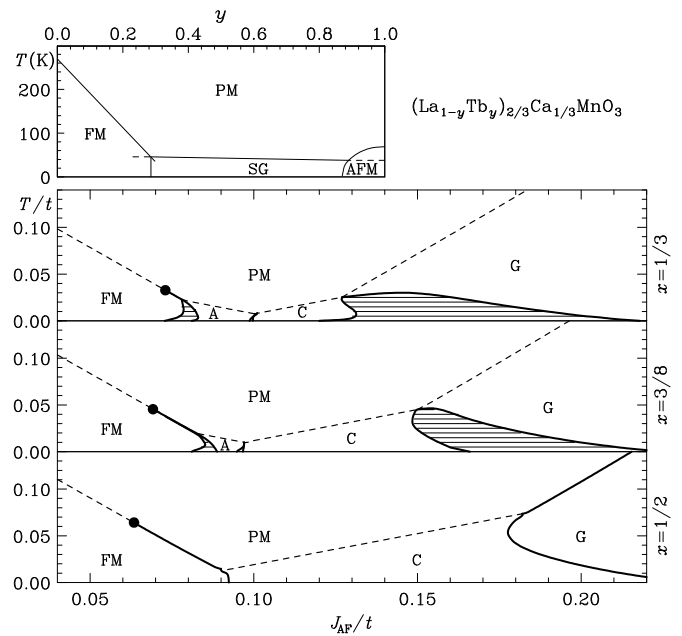,width=3.3in,angle=90}
\end{raggedleft}
\caption{Transition temperature as function of the value
of $J_{\mathrm AF} / t$ for concentrations $x=1/3$,  $x=3/8$,
and $x=1/2$. 
The dashed (solid) lines
correspond to continuous (first order) transitions and a circle has
been plotted at the tricritical PM-FM point. In the top panel we
sketch  experimental results from Ref. \protect\cite{Metal97} where
$x=1/3$. The compound
(La$_{1-y}$Pr$_y$)$_{5/8}$ Ca$_{3/8}$MnO$_3$ studied in
Ref. \protect\cite{Uetal99} has $x=3/8$.}
\label{fig:JAF}
\end{figure}

iii) At moderate to high dopings, the FM-PM transition
becomes discontinuous, if the AFM couplings are sufficiently
large. 
The onset for first order transitions at $x=1/2$ is 
$J_{\mathrm AF}/t\approx 0.06$. Unlike the previous two cases, this 
transition takes place between phases of similar electronic density.
First and second order transition lines are separated by tricritical
points.

iv) In an interval of $J_{\mathrm AF}/t$, which depends on the
doping level, we also find phase transitions between the PM
and A-AFM and C-AFM phases, that are of second order
(see Fig. \ref{fig:JAF}). At low temperatures there appear
FM, C-AFM, A-AFM, and G-AFM phases separated by first order
transitions with its associated phase separation regions,
as shown in Fig~\ref{fig:T0}.

As we see, the DEM complemented with AFM superexchange 
interactions between 
the localized spins give rise to a very rich magnetic phase
diagram that contains first and second order transitions between
phases with different magnetic order. 

In order to set a common frame for comparison with with standard
approximations~\cite{G60,AGG99}, we note the free energy of the system
is made up of an entropy term, due to the thermal fluctuations of the
core spins, an almost temperature independent contribution from the
electrons and another temperature independent term due to the direct
AFM coupling between the core spins. For instance, in the PM-FM case,
we can write: ${\cal F} = 3 J_{\mathrm AF} M^2 + E_{\mathrm elec} ( M
) -T S(M)$ where $S ( M )$ is the entropy of a spin in an effective
magnetic field producing magnetization $M$. We can expand: $S ( M ) =
-(\frac{3}{2} M^2 + \frac{9}{20} M^4 + \frac{99}{350} M^6 + ...)$ and
$E_{\mathrm elec} ( M ) = c_1 M^2 + c_2 M^4 + c_3 M^6 + ...$ where
$c_1, c_2$ and $c_3$ are functions of the band filling, and $c_1$ is
always negative ($c_1$, $c_2$ and $c_3$ are obtained fitting the numerical results for $E_{\mathrm elec}$). If there is a continuous transition, the critical
temperature is given by $T_{\mathrm C} = (2 |c_1| - 6 J_{\mathrm
AF})/3$.  The transition becomes discontinuous when the quartic term
in ${\cal F} (M)$ is negative. This happens if $c_2 < 0$ and $T < 20 /
9 | c_2 |$.  Thus, if $J_{\mathrm AF} > | c_1 | / 3 - 10 |c_2| / 9$,
and $c_2 < 0$, the transition becomes of first order. A tricritical
point appear in the transient between first and second order
transitions.

The fact that $c_2 < 0$ is due to the energetics of the electrons in the
disordered spin background. In a fully polarized system, $M=1$, the 
electrons propagate in a perfect lattice. If $M = 0$ the spins are
completely disordered, and our results reduce to those reported 
in\cite{RZB96,CVB98}. 

Standard approximations \cite{G60,AGG99} to the phase diagram of the
DEM use the virtual crystal approximation, in which the cubic density
of states is scaled by the average value $\langle {\cal T} (
\mbox{\boldmath$S$}_i,\mbox{\boldmath$S$}_j ) \rangle$, defined in
Eq.(\ref{hamil}). This approximation suffices to describe the main
features of the phase diagram when $J_{\mathrm AF} = 0$, but
overestimates the kinetic energy of the electrons moving in the
disordered spin background.  The effect is more pronounced near half
filling, when the electronic contribution is the largest, and $c_2$
is positive on the virtual crystal scheme. As our
calculation takes fully into account the propagation of the electrons
in a disordered environment, we think that the existence of a first
order PM-FM transition when $T_{\mathrm C}$ is suppressed is a robust
feature of the model.

At zero temperature, our calculation leads to a richer phase diagram 
to that calculated within the Dynamical Mean Field 
Approximation\cite{CMS00}. As mentioned in the preceding section,
our method coincides with this approximation when implemented
in a Bethe lattice. The topology of a cubic lattice allows
for the possibility of A-AFM and C-AFM phases.

\begin{figure}
\centerline{\epsfig{file=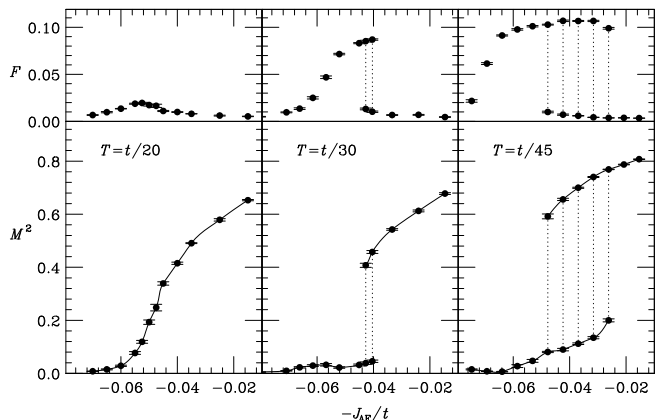,width=2.2in,angle=90}}
\caption{Monte Carlo results for the squared magnetization (bottom), and the
$\mbox{\boldmath$k$} = ( 2 \pi / 8 , 0 , 0 )$ squared Fourier component
of the  magnetization (top) in $8 \times 8 \times 8$
lattices, as function of $J_{\mathrm AF} / t$, for $x=1/2$
at different temperatures.}
\label{fig:MC}
\end{figure}

We have developed an exact Monte Carlo algorithm to study the
DEM. This approach is based in a Path Integral formulation that allows
to simulate on lattices much larger than in an usual Hamiltonian
formulation. Full details of the method will be given elsewhere
\cite{OURHMC}.  The first data of the Monte Carlo computation confirm
the robustness of the present results. Simulations in the parameter
region depicted in Fig.~\ref{fig:JAF} show a very clear evidence for a
first order transition in lattices up to $12 \times 12 \times 12$
sizes. In Fig.~\ref{fig:MC} we show data on a $L=8$ lattice at
half-filling at several temperatures. Note the large region of
metaestability marked by the vertical lines.  It is also clear that
fluctuations lower the transition temperatures from their mean-field values, 
as it also
happens in the three-dimensional Heisenberg model~\cite{O3NOSOTROS}.  In
addition, we find a helicoidal spin structure at sufficiently low
temperatures, which replaces, partially, the A-AFM and C-AFM phases
discussed earlier.

Turning again to the Mean-Field approach, let us recall that while a
continuous transition is changed into a smooth crossover in an applied
field, a first order transition survives until a critical field is
reached. The transition takes place between two phases with finite,
but different, magnetization, in a similar way to the liquid-vapor
transition. The PM-FM line of first order transitions for dopings
close to $x=0.5$ ends in a critical point, $( T_{\mathrm c} ,
H_{\mathrm c} )$.  For $J_{\mathrm AF} = 0.08 t$, the critical field
varies from $H_{\mathrm c} = 0.00075 t \approx 2.2$T at $x = 0.5$ to
$H_{\mathrm c} = 0.0002 t \approx 0.6$T at $x = 0.3$, while
$T_{\mathrm c} \approx T_{\mathrm C}$ and $T_{\mathrm C}$ is the Curie
temperature at zero field, shown in Fig.~\ref{fig:JAF}.

\section{Conclusions.}

We have shown that the phase diagram of double exchange systems
is richer than previously anticipated, and differs substantially from
that of more conventional itinerant ferromagnets. We have described 
first order transitions which are either intrinsic to the double exchange 
mechanism, or driven by the competition between it and AFM
couplings. In particular, we find that, in the doping range relevant
for CMR effects, AFM interactions of reasonable
magnitude change the PM-FM transition from continuous to first
order. The existence of such a transition has been argued, on phenomenological
grounds, in order to explain the observed data in a variety
(but not all) of doped manganites 
in the filling range $x \sim 0.3 - 0.5$\cite{Hetal99,Metal99c}.
The generic phase diagram that we obtain is consistent with 
a number of observations:

i) Materials with a high transition temperature (low AFM 
couplings) have a continuous PM-FM transition, with no evidence for 
inhomogeneities or hysteretic effects. The paramagnetic state shows metallic
behavior.

ii) The PM-FM transition in materials with low transition temperature
(significant AFM couplings)
is discontinuous. Near $T_{\mathrm C}$ inhomogeneities and hysteretic behavior
are observed. The transport properties in the paramagnetic phase
are anomalous.

iii) Substitution of a trivalent rare earth for another one with
smaller ionic radius (i.e., compositional changes that do not modify
the doping level) diminishes the ${\mathrm Mn-O-Mn}$ bond angle,
reducing the conduction bandwidth, $W=12 t$
\cite{Getal96,FoLaOb99}. Assuming that the AFM coupling, $J_{\mathrm
AF}$, does not change significantly, the ratio $J_{\mathrm AF}/t$
increases; therefore, the doping level $y$ in series of the type
$({\mathrm RE}_{1-y}{\mathrm RE}_y)_{1-x}{\mathrm AE}_x {\mathrm Mn
O}_3$ might be traded by $J_{\mathrm AF}/t$. The top panel of
Fig.~\ref{fig:JAF} shows the experimental magnetic phase diagram of
(La$_{1-y}$ Tb$_y$)$_{2/3}$ Ca$_{1/3}$ MnO$_3$, as taken from Ref.
\cite{Metal97}.  We note the similarities with the phase diagrams of
the DEM in the plane $(J_{\mathrm AF}/t,T/t)$ at fixed $x$. The phases
A-AFM and C-AFM at intermediate $J_{\mathrm AF}/t$ could become spin
glass like phases in presence of disorder.

iv)  The first order PM-FM transition reported here survives in the
presence of an applied field. A critical field is required
to suppress it (hysteretic effects in an applied field have been
reported in \cite{Betal99}).

\section{Acknowledgements.}
We are thankful for helpful conversations to L. Brey, 
J. Fontcuberta,
G. G\'omez-Santos, C. Simon, J.M. De Teresa, and especially to R. Ibarra. 
V. M.-M. is MEC fellow.
We acknowledge financial support from grants PB96-0875,
AEN97-1680, AEN97-1693, AEN99-0990 (MEC, Spain) and (07N/0045/98) (C. Madrid).

\end{document}